\documentclass[%
 reprint,
 superscriptaddress,
 amsmath,amssymb,
 aps,
prapplied,
floatfix,
]{revtex4-2}
\usepackage{graphicx}
\usepackage{dcolumn}
\usepackage{bm}
\usepackage{siunitx}
\usepackage[columnwise]{lineno}

\usepackage{hyperref}

\begin{document}
\preprint{}
\title{Toward Robust Autotuning of Noisy Quantum Dot Devices}

\author{Joshua Ziegler}
\email{joshua.ziegler@nist.gov}
\affiliation{National Institute of Standards and Technology, Gaithersburg, MD 20899, USA}

\author{Thomas McJunkin}
\affiliation{National Institute of Standards and Technology, Gaithersburg, MD 20899, USA}
\affiliation{Department of Physics, University of Wisconsin-Madison, WI 53706, USA}

\author{E. S. Joseph}
\affiliation{Department of Physics, University of Wisconsin-Madison, WI 53706, USA}

\author{Sandesh S. Kalantre}
\affiliation{Joint Quantum Institute, University of Maryland, College Park, MD 20742, USA}
\affiliation{Joint Center for Quantum Information and Computer Science, University of Maryland, College Park, MD 20742, USA}

\author{Benjamin Harpt}
\affiliation{Department of Physics, University of Wisconsin-Madison, WI 53706, USA}

\author{D. E. Savage}
\affiliation{Department of Materials Science and Engineering, University of Wisconsin-Madison, WI 53706, USA}

\author{M. G. Lagally}
\affiliation{Department of Materials Science and Engineering, University of Wisconsin-Madison, WI 53706, USA}

\author{M. A. Eriksson}
\affiliation{Department of Physics, University of Wisconsin-Madison, WI 53706, USA}

\author{Jacob M. Taylor}
\affiliation{National Institute of Standards and Technology, Gaithersburg, MD 20899, USA}
\affiliation{Joint Quantum Institute, University of Maryland, College Park, MD 20742, USA}
\affiliation{Joint Center for Quantum Information and Computer Science, University of Maryland, College Park, MD 20742, USA}

\author{Justyna P. Zwolak}
\email{jpzwolak@nist.gov}
\affiliation{National Institute of Standards and Technology, Gaithersburg, MD 20899, USA}

\date{\today}
\begin{abstract}
The current autotuning approaches for quantum dot (QD) devices, while showing some success, lack an assessment of data reliability. 
This leads to unexpected failures when noisy or otherwise low-quality data is processed by an autonomous system.
In this work, we propose a framework for robust autotuning of QD devices that combines a machine learning (ML) state classifier with a data quality control module.
The data quality control module acts as a ``gatekeeper'' system, ensuring that only reliable data are processed by the state classifier.
Lower data quality results in either device recalibration or termination. 
To train both ML systems, we enhance the QD simulation by incorporating synthetic noise typical of QD experiments.
We confirm that the inclusion of synthetic noise in the training of the state classifier significantly improves the performance, resulting in an accuracy of $95.0(9)\,\%$ when tested on experimental data.
We then validate the functionality of the data quality control module by showing that the state classifier performance deteriorates with decreasing data quality, as expected.
Our results establish a robust and flexible ML framework for autonomous tuning of noisy QD devices.
\end{abstract}

\maketitle
\section{Introduction}
Gate-defined semiconductor quantum dots (QDs) are a quantum computing technology that has potential for scalability due to their small device footprint, operation at few Kelvin temperatures~\cite{petit_universal_2020, yang_operation_2020}, and fabrication with scalable techniques~\cite{phystoday_qd_vandersypen_2019, PhysRevApplied.14.024066_chanrion2020, zwerver2021qubits}.
However, minute fabrication inconsistencies present in current devices mean that every qubit must be individually calibrated or tuned~\cite{vandersypen_interfacing_2017,phystoday_qd_vandersypen_2019}.
To enable more efficient scaling, this requirement must be met with automated methods.

\begin{figure*}[t]
\centering
\includegraphics{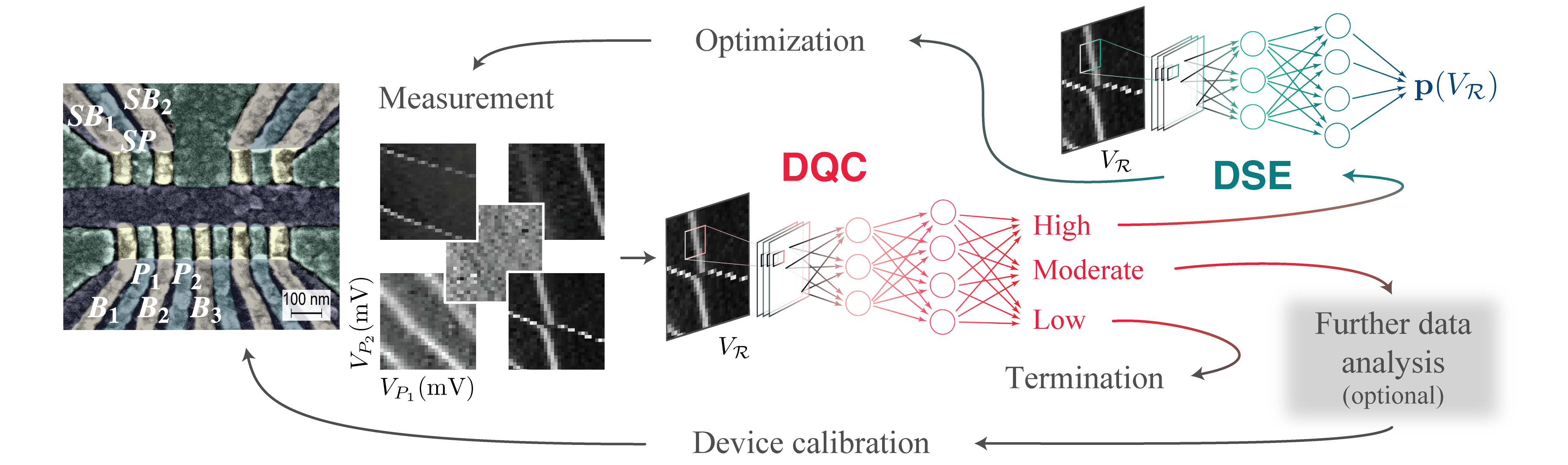}
\caption{\label{fig:framework} 
Framework for QDs autotuning with data quality assessment.
A false-color scanning electron micrograph of a Si/Si$_{x}$Ge$_{1-x}$ quadruple QD device. 
The gates in the upper channel (barriers $SB_1$, $SB_2$, and a plunger $SP_1$) are used to form a charge sensor for the QDs formed in the lower channel (using barriers $B_1$, $B_2$, and $B_3$, and plungers $P_1$ and $P_2$).
There are two consecutive machine learning modules guiding the autotuning system: DQC is used to determine the quality of the measured scan and DSE is used to assess the state of the device. 
The autotuning loop begins with the QD device shown on the left.
A two-dimensional voltage sweep of two plunger gates $(V_{P_1},V_{P_2})$ is measured by a QD charge sensor in the upper left channel. 
The numerical gradients of the measurements are then fed into the DQC module to determine whether the scan is suitable for classification. 
Depending on the returned quality class, the scan is passed to the DSE module for state assessment and optimization (the high-quality class), the device is recalibrated to improve the data quality (the moderate-quality class), or the autotuning loop is terminated (the low-quality class). 
Before recalibration or termination, further data analysis could be performed to better guide the recalibration.}
\end{figure*}

Recently, many advances have been made toward automated calibration of QD devices
~\cite{Baart2016, Mills19-CAT, Moon2020, Zwolak20-AQD}.
Automated methods have been used to tackle many stages of the calibration process, from understanding fabrication results~\cite{mei2021-OQF} and coarse device tune-up~\cite{Baart2016, Kalantre17-MLD, Zwolak2019, Moon2020, Lapointe-Major2020, Zwolak20-AQD,  Darulova2020}, to fine calibrations of device parameters~\cite{vanDiepen2018, Mills19-CAT}.
The techniques used for automation follow two main schools of thought: script-based algorithms and machine learning (ML) methods.
While appealingly simple, methods that rely on conventional algorithms are susceptible to noise and transfer poorly across devices~\cite{Lapointe-Major2020}.
On the other hand, methods that rely on ML algorithms have the flexibility to avoid being confounded by noise if provided with proper training data~\cite{Darulova2020, nguyendeep2021}, but require large labeled datasets for training and lack information on the reliability of the ML prediction.

Automated tuners, both ML- and non-ML-based, make many sequential decisions based on limited data acquired at each step.
In such a framework, small error rates can quite rapidly compound into high failure rates~\cite{Durrer19-ATQ}.
One key failure mode of QD autotuning algorithms is signal-to-noise ratio (SNR) reductions during the tuning process~\cite{Durrer19-ATQ,Lapointe-Major2020, Darulova2020}.
One way to avoid tuning failure and to promote trust in ML-based automation~\cite{Stanton2021} is to develop assessment techniques to verify the quality of data before moving forward with tuning.

In this manuscript, we present a framework for robust automated tuning of QD devices that combines a convolutional neural network (CNN) for device state estimation with a CNN for assessing the data quality, similar to approaches for general image noise estimation~\cite{xu_fast_2020}.
Inspired by recent efforts on using physics-based data augmentation to improve training of ML models~\cite{crosskey_physics-based_2018, Gomez_Gonzalez_2018, darulova2020evaluation, Lui_phyaug_2021}, we use synthetic noise characteristic of QD devices to train these two networks.
To establish the validity of the noisy dataset, we first train a CNN module to classify device states and achieve an accuracy of $95.0(9)\,\%$ on experimental data~\footnote{We use a notation value(uncertainty) to express uncertainties, for example $1.5(6)\ {\rm cm}$ would be interpreted as $(1.5\pm0.6)~{\rm cm}$. All uncertainties herein reflect the uncorrelated combination of single-standard deviation statistical and systematic uncertainties.}---an improvement of $46\,\%$ over the mean accuracy of neural networks trained on noiseless simulations. 
We then use the noisy simulations to train a data quality control module for determining whether the data is feasible for state classification.
We show that the latter not only makes intuitive predictions, but also that the predicted quality classes correlate with changes in classifier performance.
These results establish a scalable framework for robust automated tuning and manipulation of QD devices.
Furthermore, we openly publish the datasets of noisy simulated measurements (QFlow 2.0) as well as a labeled experimental dataset to further ML research in the QD domain~\cite{qf-data}.

The manuscript is organized as follows: In Sec.~\ref{sec:methods} we describe how we establish the simulated and experimental datasets. 
In Sec.~\ref{sec:results} we discuss how the noise augmentation improves state classifier performance, and demonstrate the effectiveness of the quality classifier. 
Finally, in Sec.~\ref{sec:conclusion} we summarize the results and discuss the outlook.

\section{Tuning with the data quality assessment framework}\label{sec:methods}
While a number of the recent automation proposals for QDs look promising~\cite{Durrer19-ATQ,Zwolak20-AQD,nguyendeep2021}, they all lack an assessment of the prediction reliability~\cite{Lakshminarayanan2017}. 
This largely stems from a lack of such measures for ML, though for some approaches the ``quantitative'' (i.e., assigning fractional states to images capturing transitions between states) rather than ``qualitative'' (i.e., assigning a single most dominant state to the whole image) nature of labels further complicates this issue.
Yet, the quantitative nature of prediction for intermediate regions in the state space is not only expected but might be necessary for successful operation~\cite{Zwolak20-AQD}.
In other words, a two-state prediction for a given scan should indicate that the scan captures a transition between those states, which is crucial for tuning~\cite{Zwolak20-AQD,Durrer19-ATQ}.
At the same time, if the SNR is low or in the presence of unknown fabrication defects, such a mixed prediction might instead indicate model confusion~\cite{Lakshminarayanan2017}.
In the latter case, if such confusion is not accounted for and corrected, it is likely to result in autotuning failure.

\begin{figure*}[t]
\includegraphics[width=1\textwidth]{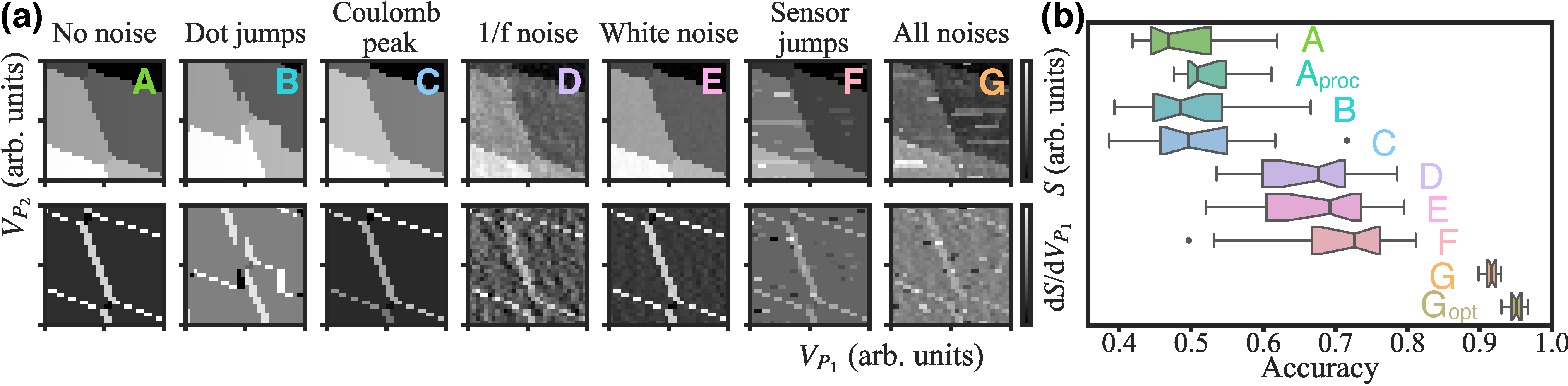}
\caption{\label{fig:samplesim_boxplot}
(a) Sample simulated charge stability diagrams as a function of plunger gates with different types of noise added. 
Top: simulated sensor ($S$) output. 
Bottom: gradient of sensor in the $V_{P_1}$ direction, d$S/$d$V_{P_1}$.
Noise magnitudes in these plots match the optimized parameters except for dot jumps (B) and the Coulomb peak (C) which are exaggerated for visibility. 
(b) Box plot showing the performance of DSE classifiers on experimental holdout dataset for models trained on: simulated noiseless data without (A) and with preprocessing ($\mathrm{A_{proc}}$), simulated data with each noise type incorporated (one at a time; plots B through F), and the optimized combination of noises (dot jumps, sensor jumps, $1/f$, and white noise; plot G).
Each box plot depicts the distribution of the performance from 20 models.
While $1/f$ noise (D), white noise (E), and sensor jumps (F) each lead to significant improvement over the model trained with noiseless simulated data (A), the optimized noise combination (G) provides a large reduction in variability as well as a significant boost in accuracy.
Optimization of the DSE model architecture further improves the performance ($\mathrm{G_{opt}}$).}
\end{figure*}

To help overcome this issue, we propose a framework where a previously introduced device state estimation (DSE) module~\cite{Kalantre17-MLD} is combined with a ML-based data quality control (DQC) module to alert the autotuning system when the measured scan is unsuitable for classification.
A diagram depicting the flow of the proposed framework is shown in Fig.~\ref{fig:framework}.
The DQC module includes a CNN classifier with a three-level output signaling the quality of a scan.
If the scan is classified as ``high quality,'' the DSE module followed by an optimization step is executed. 
For scans classified at the intermediate ``moderate quality,'' a device recalibration step is initiated.
Depending on the device and the level of system automation, this step can include readjustment of the sensor, validation of the gate cross-capacitances, or barrier gate adjustments, among other things.
To better gear the recalibration, this step could be preceded by a more detailed analysis of the image aimed at determining, e.g., the most prominent types of noise, tunnel rate issues, unintentional dots, or other issues affecting the quality of the scan.
Finally, scans with ``low quality'' indicate that there might be a bigger underlying issue.
This class results in autotuning termination.

As shown in Ref.~\cite{zhang2017beyond,xu_fast_2020}, relatively shallow CNN-based noise estimation models can be used for some image processing and denoising tasks.
However, the ability to develop and prepare such estimators hinges on the availability of training data. 
The features compromising data quality present in QD devices can be complex and vary significantly between devices.
A reliable training dataset has to account for the different types and magnitudes of undesirable features that can be encountered experimentally. 
While full control over all factors affecting data quality is unfeasible experimentally, it can be achieved (albeit within certain limits that we discuss later ) with synthetic data.
Here, we show that incorporating different types and magnitudes of simulated physical noises into the training dataset not only allows us to establish a data quality control tool, but also significantly improves performance of a state classifier on experimental data.

\subsection{Noiseless simulations}
To establish a benchmark performance for comparison with CNN classifiers trained on synthetic noise, we use a dataset of about $1.6\times10^4$ simulated noiseless measurements.
The QD simulator we use is based on a simple model of the electrical gates and a self-consistent potential calculation and capacitance model to determine the stable charge configuration~\cite{Kalantre17-MLD}. 
This simulator is capable of generating current maps and charge stability diagrams as a function of various gate voltages that reproduce the qualitative features of experimental charge stability diagrams~\cite{Zwolak2019}.
The simulated data represent an idealized device in which the charge state is sensed with perfect accuracy.
It also assumes the system is always in the ground state which results in infinitely sharp transitions.
Panel A in Fig.~\ref{fig:samplesim_boxplot}(a) shows a sample noiseless simulated stability diagram.

\subsection{Experimental data}
\label{ssec:exp_dat}
To validate the synthetic noise and test the performance of the ML modules, we establish a dataset of 756 manually labeled experimental images.
These data are acquired using two quadruple QD devices, both fabricated on a Si/Si$_x$Ge$_{1-x}$ heterostructure in an accumulation-style design with overlapping aluminum gates architecture~\cite{Angus:2007p845,Zajac:2016p054013,Dodson:2020p505001,McJunkin21-VSQ} and operated in a double dot configuration.
The gate-defined QD devices use electric potentials defined by metallic gates to trap single electrons either in one central potential, or potentials on the left and right sides of the device.
Changes in the charge state are sensed by a single electron transistor (SET) charge sensor. The charge states of the device correspond to the presence and relative locations of trapped electrons: no dot (ND), single left, central or right dot (LD, CD, RD, respectively), and double dot (DD).

Here we use experimental data from Ref.~\cite{Zwolak20-AQD}, consisting of two different datasets of 82 and 503 images, respectively, as well as data collected from a second device from a different fabrication run~\cite{McJunkin:2021R29}, resulting in 171 images.
For optimizing the synthetic noise parameters, we use randomly selected data from the first device: 80 images from the first dataset and 134 from the second dataset.
The remaining images from the first device as well as all data from the second device comprise the holdout set used for testing the trained DSE models.
The full experimental dataset is used to test the DQC module.

All images are manually labeled by two team members and any conflicting labels are reconciled through discussions with the researcher responsible for data collection.
The resulting dataset is available via the National Institute of Standards and Technology (NIST) Science Data Portal~\cite{qf-data} and at data.gov.

\subsection{Toward realistic simulations}
There are multiple sources of noise in experimental data: dangling bonds at interfaces or defects in oxides lead to noise at the device level; thermal noise, shot noise, and defects in electronics throughout the readout chain result in noise at the readout level~\cite{Connors:2019p165305,Stewart:2016p187,Spruijtenburg:2018p143001,Peters1999, Motchenbacher_Connelly_Motchenbacher_1993, Culcer_2009}. 
In many QD devices, changes in the device state are sensed by conductance shifts in a SET due to their sensitivity to transitions with no change in net charge.
The response of a SET is nonlinear, which causes variation in the signal of charge transitions.
The various types of noise manifest themselves in the measurement though distortion that might obscure or deform the features indicating the state of the device (borders between stable charge regions).

To prepare a dataset for the DQC module, we extend the QD simulator to incorporate the most common sources of experimental noise.
We consider five types of noise: dot jumps, Coulomb peak effects, white noise, $1/f$ (pink) noise, and sensor jumps. 
Experimentally, white noise, $1/f$ noise, and sensor and dot jumps appear due to different electronic fluctuations affecting a SET charge sensor. 
White noise can be attributed to thermal and shot noise while the $1/f$ noise can have contributions from various dynamic defects in the device and readout circuit~\cite{Connors:2019p165305, Paladino_2014, Motchenbacher_Connelly_Motchenbacher_1993, Culcer_2009}.
Previously, we modeled the charge sensor with a linear response, though in reality it has a nonlinear response due to the shape of the Coulomb blockade peak.
We account for this with a simple model of a SET in the weak coupling regime~\cite{PhysRevB.44.1646}.
Physically, dot jumps and sensor jumps are two manifestations of the same process: electrons populating and depopulating charge traps in the device, which we model as two-level systems with characteristic excited and ground state lifetimes. 
Dot jumps are the effect of these fluctuations on the quantum dot, while sensor jumps are the effect on the SET charge sensor.
We provide additional details on how we implement these synthetic noises in Appendix~\ref{si:noise}.
While other factors might contribute to compromising data quality, we do not consider them in this work.
However, as we show in Sec.~\ref{ssec:val}, the noise types presented here are sufficient for identifying regions within large scans that are compromised due to factors other than just noise as moderate-quality.

Each of the modeled noises can obscure or mimic charge transition line features, potentially confusing ML models. 
White noise and $1/f$ noise both generate high-frequency components that can be picked up in the charge sensor gradient. 
Additionally, the $1/f$ noise can generate shapes that look similar to charge transition lines.
Sensor jumps cause large gradients where they occur. 
Movement of the SET Coulomb peak can reduce the visibility of charge transitions if it moves to a point off the sloped sides with lower gradient and thus lower sensitivity.
Finally, dot jumps can distort the shapes of charge transition lines. 
Panels B--F in Fig.~\ref{fig:samplesim_boxplot}(a) show charge stability diagrams with each of the discussed noise types added (one at a time).

For each type of noise, we generate a distinct dataset of about $1.6\times10^4$ simulated measurements using the same device parameters as used for the simulated noiseless dataset.
To determine simulated noise parameters, we first seek to produce images qualitatively similar to reasonably noisy experimental data.
We then optimize those parameters through a semistructured grid search over a range centered at the initial value levels.
At each step, the correlation between the noise level and DSE performance on a subset of experimental images is used to guide the search.
The dataset used to train models for each noise type are generated by varying each noise parameter with a standard deviation of $1\,\%$ of the parameters' value.
Panel G in Fig.~\ref{fig:samplesim_boxplot}(a) shows a sample image with the optimized combination of noises.

The final noisy simulated dataset has $1.15 \times 10^5$ images generated by fixing the relative magnitudes of white noise, $1/f$ noise, and sensor jumps and varying the magnitudes together in a normal distribution.
The means of the magnitudes are set to 1.5 times the optimized values (to ensure that low-quality data are included in the training dataset) and the standard deviation is one third of each magnitude's value.
Fixing the relative magnitudes and varying them together allows this distribution of noise levels to approximate a range of SNR encountered in experiments.
This dataset is also available via the NIST Science Data Portal~\cite{qf-data} and at data.gov.

\subsection{Assessing data quality}
In the second phase, we focus on the development of the DQC module.
As we already stressed, the QD state labels are quantitative, so a mixed label indicates an intermediate state rather than confusion and is important for the autotuning system proposed in Ref.~\cite{Zwolak20-AQD}. 
This means that a simple entropy of a model's prediction cannot be used as a measure of confusion.
Rather, an alternative quality measure needs to be established.
To achieve this, we leverage the simulated noise framework established in the previous section to perform a controlled analysis of the DSE module performance as noise levels are varied.

\begin{figure*}
\includegraphics[width=1.\textwidth]{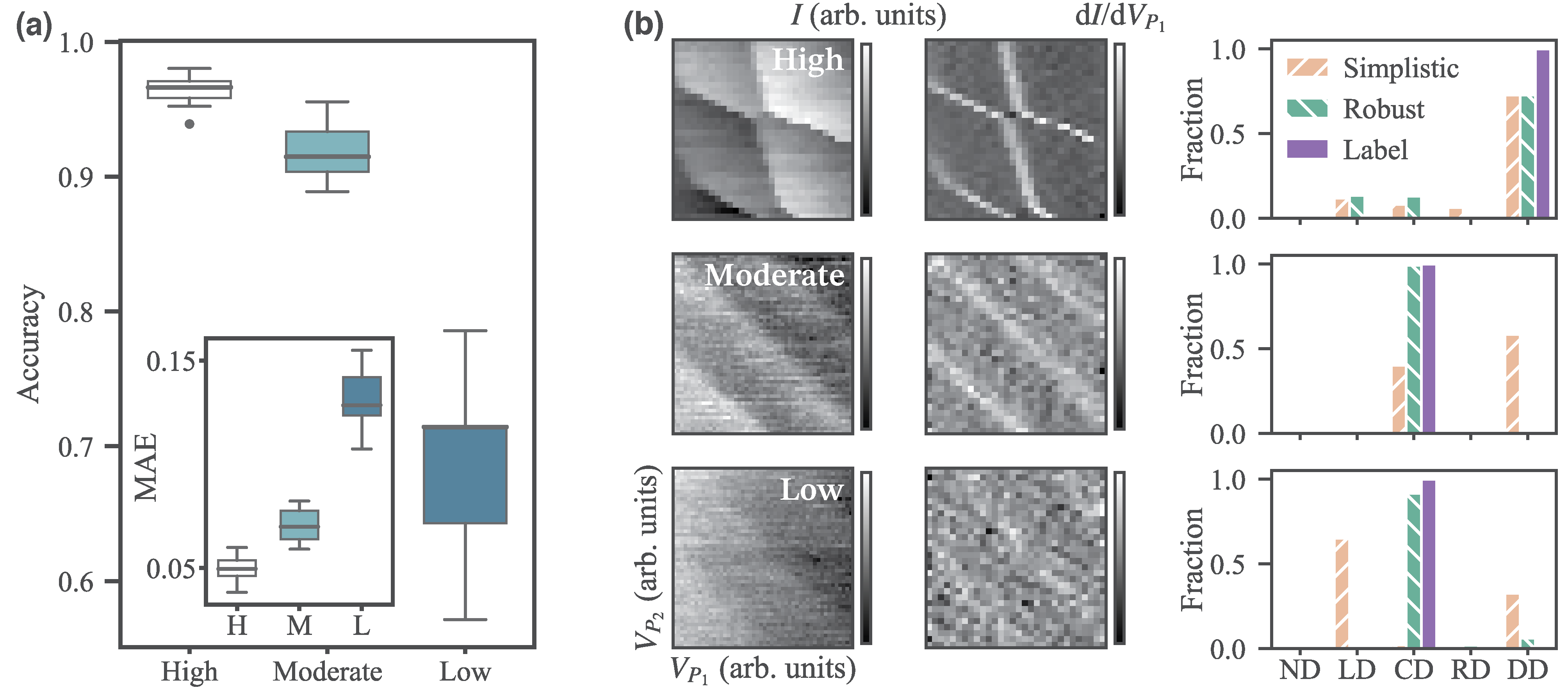}
\caption{\label{fig:exampleimages_noiseboxplot}
(a) Box plots of model accuracy for each assigned quality class for the experimental data. 
Inset: box plots of the mean absolute error (MAE) for each quality class. 
(b) Example data and predictions of both the simplistic (i.e., trained on noiseless simulated data) and robust (i.e., trained on noisy simulated data) models. 
Raw sensor data (left), gradient data (middle), and predictions (right). 
We show a high-quality DD example, a moderate-quality CD example, and a low-quality CD example. 
For the bar plot, we include the full prediction vector for the simplistic and robust models, as well as the assigned label for each image.}
\end{figure*}

In the framework presented in Fig.~\ref{fig:framework}, we propose to use three levels of data quality---high, moderate, and low---to determine the subsequent actions.
To determine the threshold between the three quality classes, we generate a dataset of $1.15\times10^5$ simulated images with varying amounts of noise added.
Since features defining the QD states are affected in distinct ways by the noise, the performance versus noise level analysis is carried out separately for each state rather than for the whole dataset (see Appendix~\ref{si:dqt} for more details).
We vary the magnitudes of all noises that negatively affect the SNR (sensor jumps, $1/f$, and white noise) together from $0$ to $7$ times the optimized noise magnitudes while keeping the dot jumps noise variation within the $1\,\%$ used previously.
This distribution of noise includes a large variation of noise levels from near-perfect data to data that has nearly no recognizable QD features.
This is necessary for establishing noise thresholds for the data quality classes that ensure saturation of the performance of the state classifier at both the low and high levels.

By evaluating a state classifier, trained on a dataset with all synthetic noises added (box plot G in Fig.~\ref{fig:exampleimages_noiseboxplot}(b)), on this dataset we determine the relationship between the noise level and performance within each QD state class.
From the correlations between noise level and performance, we establish per-QD state data quality thresholds.
The thresholds are chosen to ensure high performance of the state classifier for the high-quality data, an expected degradation of performance for data with moderate quality, and poor performance on data with low quality.
Specifically, we set the cutoffs using the relationship between the model's mean absolute error (MAE) and noise level (see Fig.~\ref{fig:SI_MAEvsNoise} in Appendix~\ref{si:noise}).

We set these cutoff levels at relatively conservative amounts of noise, which would enable a fairly risk-averse tuning algorithm.
This parameter choice could be adjusted to the needs of a given application depending on the error sensitivity of an autotuning method.
To ensure that images in the high-quality class are very reliably identified, we set the threshold between high- and moderate-quality classes to be at the noise level where the average MAE has gone up by $2.5\,\%$ of the full range, which is similar to a 2 sigma cutoff for the lower tail of a normal distribution.
We set the threshold between moderate and low quality where the average MAE has reached $50~\%$ of its full range, that is where the model is roughly equally likely to be wrong as right for a single state image.

With these thresholds, state labels, and the known amount of noise added, we then assign the simulated data with quality classes for DQC module training.
For this training, we use a distinct dataset with the same distribution of noise used to set quality class thresholds.
This dataset is also available via the NIST Science Data Portal~\cite{qf-data} and at data.gov.

\section{Results}\label{sec:results}
To prepare the data quality control module (DQC in Fig.~\ref{fig:framework}), we validate the simulated noise by training a CNN-based classifier to recognize the state of QD devices from charge stability diagrams (module DSE in Fig.~\ref{fig:framework}).
We show how each of the added noises affects the classification accuracy on a holdout subset of experimental data (see Sec.~\ref{ssec:exp_dat}) and confirm that their combination leads to significant improvement in performance, suggesting increased similarity between the simulated and experimental data.
We then use the noisy simulated data to train the DQC module. 
The full experimental dataset is used to confirm the correlation between the predicted quality class and classification performance. 
Finally, we use large scans to show that the optimized model (called ``robust'') outperforms the model trained on noiseless data (called ``simplistic'') and show how the predicted quality classes overlap with the confusion of the DSE module.

\subsection{Robust state classification}
To determine how the considered noise types affect the performance of the DSE classifier, we modify the simulation with each type of noise individually and evaluate models trained with that data on the experimental holdout dataset.
For initial testing, we optimize a CNN architecture defining the simplistic model used for state recognition on simulated noiseless data using the Keras Tuner API~\cite{omalley2019kerastuner} (see Appendix~\ref{si:ml} for additional information). 

Figure~\ref{fig:samplesim_boxplot}(b) summarizes the results of these tests. 
As a benchmark, we include the $48.7(5.5)\,\%$ test accuracy for models trained on simulated data without noise added (box plot A in Fig.\ref{fig:samplesim_boxplot}(b)).
As expected, the high validation accuracy of $93.6(9)\,\%$ achieved during training drops significantly when the models are tested on experimental images. 
Previous work suggests that some data processing techniques used to help suppress experimental noise might help with the performance~\cite{Zwolak20-AQD}.
Our analysis confirms that preprocessing of experimental data, as suggested in Ref.~\cite{Zwolak20-AQD}, improves the average accuracy and reduces the variance between models.
However, the observed accuracy of $51.9(3.6)\,\%$ (box plot A$_{\mathrm{proc}}$) on the experimental holdout dataset is still much lower than necessary for reliable state assessment.

\begin{figure*}
\includegraphics{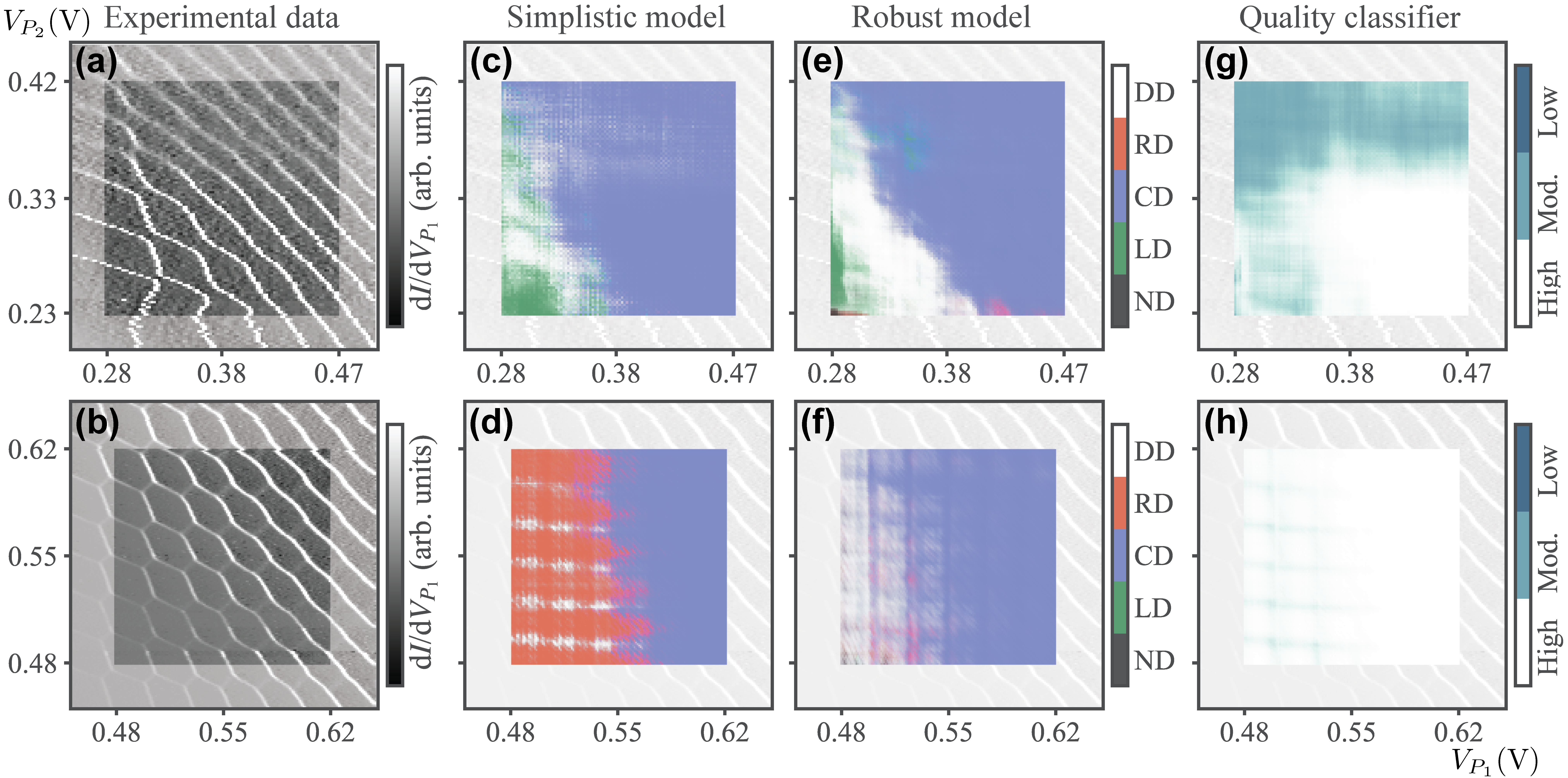}
\caption{\label{fig:eval_maps}
(a, b) Full charge stability diagrams of two double QD devices.
In (a), a few characteristic noises can be seen: minor $1/f$ or white noise is seen in the speckling throughout, jumps in the transition due to slow tunnel rates at the bottom of the image, and smearing of the transitions near the top of the image due to fast tunnel rates.
Visualization of the prediction of an average simplistic state classifier (i.e., model trained on noiseless simulated data) (c, d), and the optimized robust state classifier (i.e., model trained on noise-augmented simulations) (e, f). 
The color at each point is the average of the color of each state weighted by the model's prediction.
Hue is averaged by angle in hue space, e.g., blue and green are averaged to teal. 
(g, h) Visualization of the predictions of the DQC module.
}
\end{figure*}

When looking at the various types of noise individually, our analysis reveals that $1/f$ noise (box plot D in Fig.~\ref{fig:samplesim_boxplot}(b)), white noise (box plot E), and sensor jumps (box plot F) most significantly improve the models' performance, with $66.4(7.8)\,\%$, $66.6(8.7)\,\%$, and $70.4(8.7))\,\%$ accuracy, respectively.
Dot jumps (box plot B) and Coulomb peaks (box plot C) turn out to be unhelpful on their own. 
The former seems to affect the performance negatively.
Combining all types of noise results in a significant improvement in both the performance and variance of the resulting models, with an accuracy of $91.6(8)\,\%$ (box plot G in Fig.~\ref{fig:samplesim_boxplot}(b)).
For comparison, in the context of simulated transport data, previous work found that only the sensor jumps, $1/f$, and white noise improved classifier performance, though the observed improvements were not significant~\cite{darulova2020evaluation}.
We note that, when combining these noises, we use a varied SNR (by varying sensor jumps, $1/f$, and white noise together).
This uniformly tunes the SNR between simulated images as a replacement for the explicit Coulomb peak.
Effectively, this results in a varying visibility of charge transition lines but with more uniformity.

Finally, since the models' architecture we use is optimized for a noiseless dataset, we reoptimize the CNN architecture using the noisy simulated dataset.
This allows us to find a model that is structurally best suited to that type of data and thus further improves the performance.
With these changes, we find an increase in the classification accuracy by about $3.4\,\%$ to $95.0(9)\,\%$ (box plot G$_{\mathrm{opt}}$ in Fig.~\ref{fig:samplesim_boxplot}(b)).
We also test preprocessing of the data to remove extreme values for completeness and find no significant difference at $95.8(1.2)\,\%$ accuracy.
Additional information about the datasets and model architectures used in Fig.~\ref{fig:samplesim_boxplot}(b) are provided in Table~\ref{tab:datasetarchitecture} in Appendix~\ref{si:ml}.
Comparing box plots A$_{\mathrm{proc}}$ and G$_{\mathrm{opt}}$ shows the high level of improvement in QD state classification we are able to achieve by adding noise to the simulated training set and optimizing the model.

\subsection{Data quality control system}
The purpose of the DQC module is to filter data that would likely be unsuitable for the DSE module.
Identifying the specific issues affecting data quality is left to the (optional) ``further data analysis'' module which is not part of this work. 
However, we find that even though we use noise-enhanced data, the DQC module correctly flags regions affected by other issues, such as incorrectly set tunnel rates.

To confirm the validity of the thresholds used to define the three quality classes, we use the full experimental dataset. 
The DQC module applied to the experimental images classified 607 images as high quality, 135 images as moderate quality, and 14 images as low quality. 
Figure~\ref{fig:exampleimages_noiseboxplot}(a) shows the performance of the 20 optimized state classifiers (shown in box plot G$\mathrm{_{opt}}$ in Fig.~\ref{fig:samplesim_boxplot}(b)) for each quality class.
The error bars represent the variation in performance between the 20 models.
The DSE module performs well on data classified as high quality, with $96.4(9)\,\%$ prediction accuracy, and begins to decrease for the moderate class at $91.9(2.1)\,\%$. 
For data in the low-quality class the models' performance decreases to $69.3(5.6)\,\%$.
The variance in performance also increases as the data quality degrades.
To account for the expected partial predictions between QD states, we further validate this correlation using a fine-grained metric.  
We use the MAE to capture elementwise deviation.
The inset in Fig.~\ref{fig:exampleimages_noiseboxplot}(a) shows the MAE between the assigned and predicted labels for the three quality classes.
The observed correlations in accuracy with the quality class are also seen in MAE.
This analysis confirms that the moderate-quality class does indeed capture reductions in SNR that mildly affect model performance, while the low-quality class identifies images that are substantially more difficult for the DSE module.

Figure~\ref{fig:exampleimages_noiseboxplot}(b) shows sample experimental images from each of the quality classes and bar plots of the state prediction vectors for the simplistic and robust state classifiers, as well as the manually assigned labels. 
The top row shows a high-quality DD example correctly classified by both models, as indicated by the largest DD component in the bar plot.
The middle row shows a sample CD image assessed to have moderate quality and the bottom row shows a low-quality CD image. 
Both moderate- and low-quality images are incorrectly classified by the simplistic model. 
The quality of the bottom image in  Fig~\ref{fig:exampleimages_noiseboxplot}(b) makes it hard for a human to identify the state.
Here, the simplistic model is confused between LD and DD states, while the robust model correctly identifies this image as CD.
This illustrates the level of improvement that noisy training data provides for our DSE module.

\subsection{Validating autotuning framework}
\label{ssec:val}
Finally, we assess the viability of the proposed framework by performing tests of the DSE and DQC modules over two large experimental scans shown in Figs.~\ref{fig:eval_maps}(a) and \ref{fig:eval_maps}(b).
Figure~\ref{fig:eval_maps} shows comparisons of classification performance between sample models trained on noiseless (c), (d) and noisy (e), (f) simulated data along with the predicted quality class (g), (h).

We use a series of 60~\si{\milli\volt} by 60~\si{\milli\volt} scans sampled at every pixel~\footnote{One pixel corresponds to 2~\si{\milli\volt} in Fig.~\ref{fig:eval_maps}(a) and 1~\si{\milli\volt} in Fig.~\ref{fig:eval_maps}(b).} within the large scans and leaving a 30~\si{\milli\volt} margin at the boundary to ensure that each sampled scan is within the full scan boundaries.
From Figs.~\ref{fig:eval_maps}(c) and \ref{fig:eval_maps}(d) we see that the simplistic model does fairly well on the parts of scans where the SNR is good, but it becomes less reliable when the charge transitions are less clear.
In the first scan, this is manifested by random speckling of the DD prediction within the CD region (the top half of the scan) as well as by the frequent changes in state assessment for images sampled within a couple of pixels (the left half of that scan).
A similar effect is visible in the left half of the second scan, where the prediction oscillates between RD and DD.
For comparison, the predictions of the robust model, shown in Figs.~\ref{fig:eval_maps}(e) and \ref{fig:eval_maps}(f), are much more stable and accurate.

While areas with mixed labels are produced by both models, for the robust model, they are primarily indicative of transitions between states.
For the simplistic model, mixed labels are assigned also within single-state parts of the scans.
Such labels should not be used for autotuning as they will degrade the optimization step (see Fig.~\ref{fig:framework}).
Finally, even for the robust model there are some misclassifications in both images, particularly in the top left of Fig.~\ref{fig:eval_maps}(e) and in areas where interdot transitions are more prominent in Fig.~\ref{fig:eval_maps}(f).

However, a side-by-side comparison of panels (e) and (g) [as well as (f) and (h)] in Fig.~\ref{fig:eval_maps} reveals that regions that are misclassified by the DSE module closely match regions flagged as moderate quality by the DQC module.
This validates the DQC module as a tool to determine if the scan quality is sufficient for reliable state assessment or whether the device is in need of recalibration.
Overall, these state and data quality classification maps show that the DQC and DSE modules, when put together, provide reliable high level information for autotuning algorithms.

\section{Summary}\label{sec:conclusion}
Our results show that adding physical noise to simulated data can dramatically improve the performance of machine learning algorithms on experimental data. 
Importantly, we are able to achieve high level performance without any preprocessing or denoising of the data.
We also show how the synthetic noise can be used to develop ML tools to assess the quality of experimental data and that the assigned data quality correlates with state classifier performance, as desired.
Combining these tools enables a framework we outlined in Fig.~\ref{fig:framework}, in which the data quality control module determines whether to move forward with state classification and optimization.
This framework is an important step toward autotuning of QD devices with greater reliability.

We note that the thresholds used to establish the quality classes in the data quality control module are chosen to provide meaningful separation.
However, depending on the application's risk tolerance, these thresholds can be adjusted to obtain the error rates needed to prevent failure of an autotuning algorithm.
Beyond the classification of the data quality, our flexible synthetic noise model allows for extensions in which the data are labeled by the exact type and level of noise rather than the overall quality.
ML models can then be trained to predict the predominant types of noise, which in turn would enable tailored recalibration actions to mitigate them.

Broadly, our noise augmentation approach confirms that perturbing simulated data with realistic, physics-based noise can vastly improve the performance of simulation-trained ML models.
This may be a useful insight for other research combining ML and physics.
From a domain shift perspective, the observed performance increase could be attributed to the physical noise augmentation shifting the training data distribution nearer to the experimental test distribution~\cite{pan_survey_2010}.
Additionally, our data quality control module presents a paradigm for ML reliability estimation in which physically motivated noise models are used to determine whether to move forward with data classification.

\begin{figure*}
\includegraphics{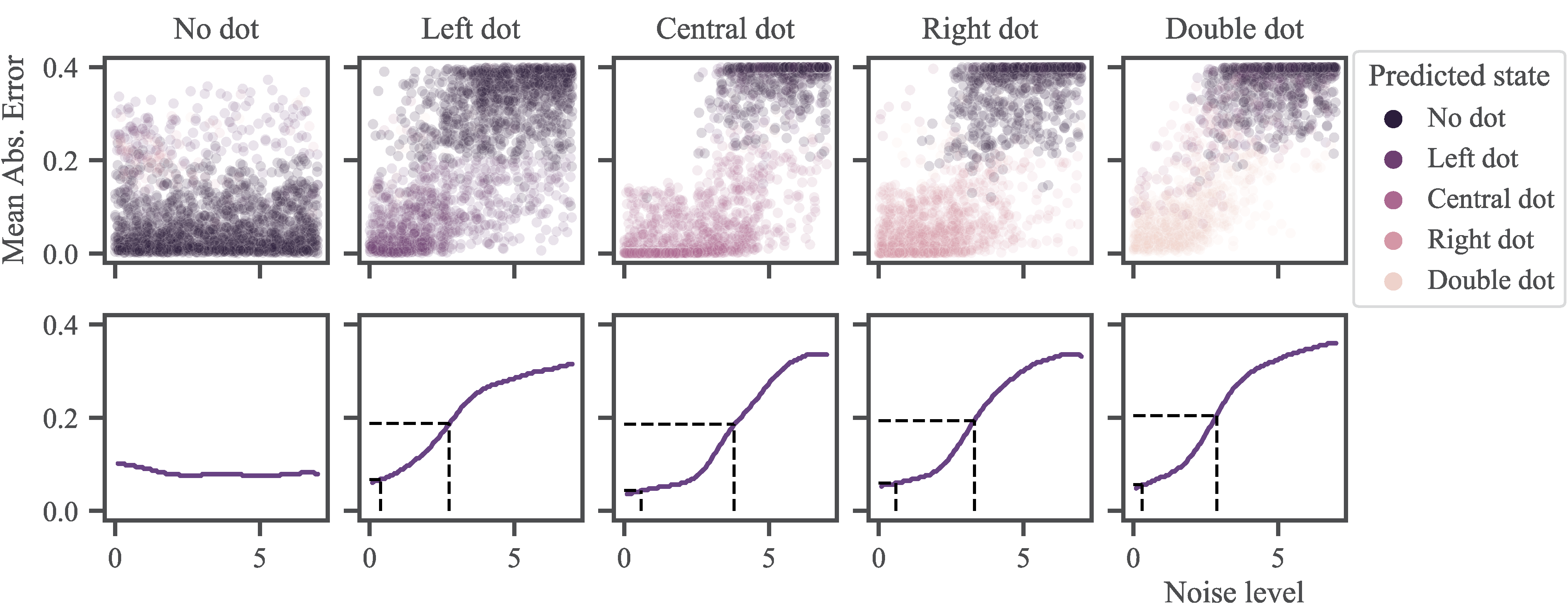}
\caption{\label{fig:SI_MAEvsNoise}
Top row: plots of the MAE of the DSE used to set noise thresholds versus the simulated noise level. 
The scatter plot is colored by the predicted state. 
Bottom row: the solid lines show the means of the MAE at each noise level. 
The dashed lines illustrate the $2.5\,\%$ and $50\,\%$ MAE levels used to set the thresholds for the DQC module.
}
\end{figure*}

\begin{acknowledgments}
This research was performed while J.Z. held a NRC Research Associateship award at the National Institute of Standards and Technology (NIST).
This research is sponsored in part by the Army Research Office (ARO), through Grant No.\ W911NF-17-1-0274. 
S.K. gratefully acknowledges support from the Joint Quantum Institute--Joint Center for Quantum Information and Computer Science Lanczos graduate program. 
We acknowledge the use of clean room facilities supported by The National Science Foundation (NSF) through the UW-Madison MRSEC (DMR-1720415) and electron beam lithography equipment acquired with support of the NSF MRI program (DMR-1625348). 
The development and maintenance of the growth facilities used for fabricating samples was supported by the Department of Energy, through Grant No.\ DE-FG02-03ER46028.
The views and conclusions contained in this paper are those of the authors and should not be interpreted as representing the official policies, either expressed or implied, of the ARO, or the U.S. Government. 
The U.S. Government is authorized to reproduce and distribute reprints for Government purposes notwithstanding any copyright noted herein. 
Any mention of commercial products is for information only; it does not imply recommendation or endorsement by NIST.
\end{acknowledgments}

\appendix
\section{Noise models details}\label{si:noise}
As discussed in the main text, we add five different types of noise to the simulated data: dot jumps, Coulomb peak effects, $1/f$ noise, white noise, and sensor jumps. 
Of these, the white noise is the simplest to implement by adding normally distributed noise with zero mean and fixed standard deviation at every pixel.
The standard deviation value is determined as part of the noise optimization process.
The $1/f$ noise is generated in Fourier space with random phase sampled uniformly over $[0, 2\pi)$ and a magnitude proportional to $1/(f_x^2+f_y^2)^{1/2}$. 
The Coulomb peak effect is applied using a simple model of a quantum dot in the weak coupling regime that yields a conductance lineshape of the form
\begin{equation}
\label{eq:coulombpeak}
G/G_{max}=\cosh^{-2}[A(V-V_{\mathrm{min}})]
\end{equation}
where $G$ is the conductance, $G_{max}$ is the peak conductance of the line, $A$ is a parameter that controls the linewidth and is determined during noise optimization, $V_{\mathrm{min}}$ is the peak center, and $V$ is the signal seen by the simulated sensor due to the quantum dots~\cite{PhysRevB.44.1646}. 
Dot jumps and sensor jumps are generated using the same underlying physics principles.
We model them as charge traps with characteristic excited and ground state lifetimes necessary for capturing or ejecting electrons. 
We achieve this by performing Bernoulli trials to determine if a jump occurs at a given pixel. 
This allows the jumps to follow a geometric distribution---the discrete analogue to an exponential distribution. 
Magnitudes of sensor jumps are drawn from a normal distribution with zero mean and fixed standard deviation determined during noise optimization.
Magnitudes of dot jumps are drawn from a Poissonian distribution with fixed rate also determined during noise optimization.

\section{Data Quality Control Thresholds}\label{si:dqt}
To provide better clarity on how we determine the noise level thresholds for training the DQC module, here we show plots of the data used to set these thresholds.
The top row in Fig.~\ref{fig:SI_MAEvsNoise} shows a series of scatter plots of the MAE between the true labels and the DSE model predictions as a function of noise level.
The model's architecture is optimized on noiseless data and the model is trained on noisy data.
This plot illustrates how the DSE performance changes as the noise level increases, revealing a roughly sigmoidal relationship.
The noise level where the MAE sharply rises varies between the LD, CD, RD, and DD states.
For the ND, state the model has on average small error regardless of the noise level.

The dashed lines in the bottom row of Fig.~\ref{fig:SI_MAEvsNoise} indicate the lower and upper thresholds at $2.5\,\%$ and $50\,\%$ of the full range of the MAE for LD, CD, RD, and DD states.
The lower threshold is fairly conservative and captures a modest rise in MAE.
At the upper threshold, on the other hand, the slope of the mean of the MAE is near its maximum and the model rapidly becomes less reliable.
These thresholds can be further adjusted based on the specific application.

Since we find no clear dependence of the MAE for ND on the noise level, the ND thresholds are set separately.
Above the $50\,\%$ thresholds, the DSE has trouble distinguishing between ND and any other state, making the ND predictions unreliable.
Thus, the upper threshold for ND is set based on the thresholds determined for the remaining four states.
The lower threshold for ND is determined in a similar fashion for consistency.

\begin{table}[b]
    \caption{Summary of datasets and model architectures.}
    \label{tab:datasetarchitecture}
    \centering
    \begin{ruledtabular}
    \begin{tabular}{lll}
        Label & Training data & Model architecture\\ \hline
        A & Noiseless & Noiseless DSE\\
        A$\mathrm{_{proc}}$ &  Noiseless, thresholded & Noiseless DSE\\  
        B & Dot jumps added & Noiseless DSE\\
        C & Coulomb peak added & Noiseless DSE\\
        D & $1/f$ noise added & Noiseless DSE\\
        E & White noise added & Noiseless DSE\\
        F & Sensor jumps added & Noiseless DSE\\
        G & All noises added & Noiseless DSE\\
        G$\mathrm{G_{opt}}$ & All noises added & Noisy DSE\\
    \end{tabular}
    \end{ruledtabular}
\end{table}

\begin{table*}[t]
    \caption{Machine learning model architectures for the noiseless DSE, noisy DSE, and DQC modules. Activation functions are either rectified linear units (ReLU) or Swish~\cite{ramachandran2017searching}}
    \label{tab:all_architectures}
    \centering
    \begin{ruledtabular}
    \begin{tabular}{llll}
        Parameter       & Noiseless DSE & Noisy DSE & DQC \\
        \hline 
        Conv. layer 1   & $(5\times5) \times 23$, stride 2 & $(7\times7) \times 22$ , stride 1 & $(7\times7) \times 184$, stride 1\\  
        Dropout layer 1 & 0.12 & 0.66 & 0.05 \\
        Layer norm.     & Yes & No & Yes \\
        Activation      & ReLU & ReLU & Swish\\
        Conv. layer 2   & $(5\times5) \times 7$, stride 2 & $(7\times7) \times 22$, stride 2 & $(3\times3) \times 249$ , stride 1\\
        Dropout layer 2 & 0.28 & 0.66 & $\dotsb$ \\
        Layer norm.     & Yes & No & Yes \\
        Activation      & ReLU & ReLU & Swish\\
        Max pool 1      & $\dotsb$ & $\dotsb$ & $(2\times2) \times 1$, stride 2\\
        Conv. layer 3   & $(5\times5) \times 18$, stride 2 & $(7\times7) \times 35$ , stride 1 & $\dotsb$ \\  
        Dropout layer 3 & 0.30 & 0.19 & $\dotsb$ \\
        Layer norm.     & Yes & No & $\dotsb$ \\
        Activation      & ReLU & ReLU & $\dotsb$ \\
        Conv. layer 4   & $\dotsb$ & $(7\times7) \times 35$, stride 2 & $\dotsb$ \\
        Dropout layer 4 & $\dotsb$ & 0.19 & $\dotsb$ \\
        Activation      & $\dotsb$ & ReLU & $\dotsb$ \\
        Ave. pool       & Yes & Yes & Yes \\
        Dense layer 1   & $\dotsb$ & $\dotsb$ & 161 \\
        Dropout layer 5 & $\dotsb$ & $\dotsb$ & 0.6 \\
        Outputs         & 5 & 5 & 3 \\
        Activation      & Softmax & Softmax & Softmax\\
        \hline
        Optimizer       & Adam & Adam & Adam\\
        Learning rate   & $3.45\times10^{-3}$ & $1.21\times10^{-3}$ & $2.65\times10^{-4}$\\
        Loss            & Cross-entropy & Cross-entropy & Cross-entropy \\
        \hline
        Trainable parameters & $7.99\times10^3$ & $1.23\times10^{5}$ & $4.63\times10^5$\\
    \end{tabular}
    \end{ruledtabular}
\end{table*}

\section{Machine Learning Model Details}\label{si:ml}
Both machine learning modules are built and trained using the TensorFlow (v.2.4.1) Keras Python API. 
We use three different model architectures: two for testing the DSE for noiseless and noisy data, and a third one in the DQC module.
All architectures are optimized to ensure high performance using the Keras Tuner~\cite{omalley2019kerastuner} and the Optuna hyperparameter tuner~\cite{optuna_2019}.
A summary of architectures and datasets used and described in Fig.~\ref{fig:samplesim_boxplot} is shown in Table~\ref{tab:datasetarchitecture}.

The optimized neural network architectures are presented in Tab.~\ref{tab:all_architectures}. 
We find from our optimization that architectures with no fully connected layers before the output layer perform better at state classification---consistent with recent results~\cite{efficientnet_tan2019}.
This is in contrast with the architecture previously used for similar tasks of quantum dot state classification~\cite{Zwolak20-AQD, Darulova2020}. 
These architectures also have up to almost three orders of magnitude less parameters compared to the original network used in Ref.~\cite{Zwolak20-AQD}.

\begin{figure*}[!hb]
\centering
\includegraphics{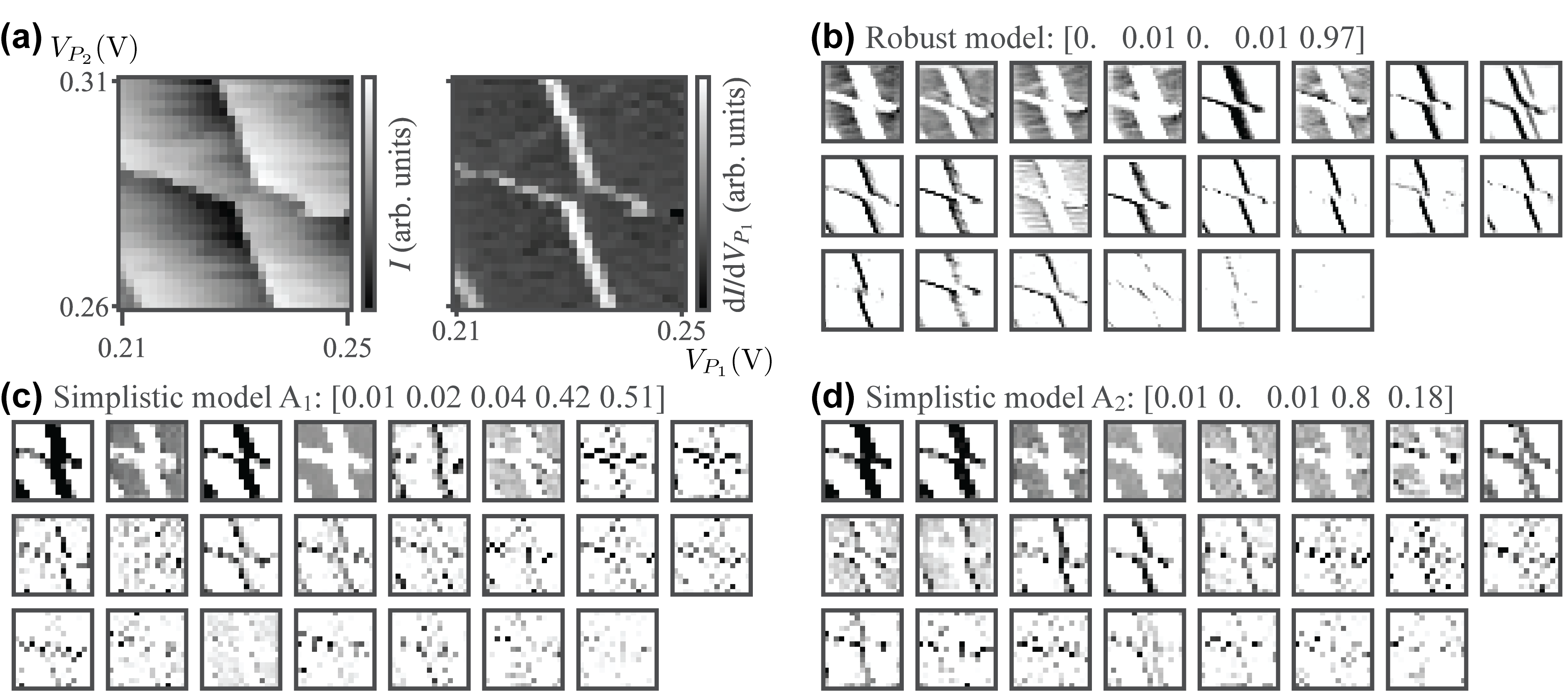}
\caption{\label{fig:SI_activations}
(a) An image from the experimental dataset.
(b) First layer outputs (after activation) of a robust (G$_{\mathrm{opt}}$) model.
(c),(d) First layer outputs of two different simplistic (A) models. 
The robust model is nearly perfectly correct, one of the simplistic models is somewhat correct, and the other simplistic model is strongly incorrect.
}
\end{figure*}

For testing the performance of our machine learning models, we train 20 models on the same simulated data (each time randomly split into training and validation).
The ML models start from a random initialization and are trained with stochastic batches of data.
These random elements can lead to different final configurations due to the nonconvex and degenerate optimization landscape. 
By training multiple models under the same conditions we can make a representative sample of models resulting from a given dataset.
An example of two different models trained under the same conditions (on $1.5 \times 10^4$ noiseless images) with different results can be seen in Fig.~\ref{fig:SI_activations}.
Here, it is likely largely by chance that the model in (c) is more correct than the model in (d) due to the noisiness in the intermediate layer outputs for both.

\onecolumngrid

\twocolumngrid

%


\end{document}